\documentclass[preprint,aps,amsmath,nofootinbib,tightenlines,floatfix]{revtex4}
\usepackage{amssymb}
\usepackage{graphicx,epsfig,amssymb}
\usepackage{multirow}
\usepackage{enumerate}
\usepackage{subfigure}

\def\met{\ensuremath{\not\!\!\!{E_{T}}}}

\def\beq{\begin{equation}}
\def\eeq{\end{equation}}
\def\bea{\begin{eqnarray}}
\def\eea{\end{eqnarray}}

\begin{document}

\preprint{UCI-HEP-TR-2011-25}

\title{ Slepton Discovery in Electroweak Cascade Decay}
\author{
         Jonathan Eckel$^{1}$\footnote{eckel@physics.arizona.edu}, 
         William Shepherd$^{2}$\footnote{shepherd.william@uci.edu}, \,and\,
        Shufang Su$^{1,2}$\footnote{shufang@physics.arizona.edu}}
        
\affiliation{
$^{1}$ Department of Physics, University of Arizona, Tucson, Arizona  85721\\
$^{2}$  Department of Physics and Astronomy, University of California, Irvine, California 92697
}

\begin{abstract}

The LHC studies on the MSSM slepton sector have mostly been focused on
direct slepton Drell-Yan pair production.  In this paper, we analyze
the case when the sleptons are lighter than heavy neutralinos and can
appear in the on-shell decay of neutralino states.   
In particular, we have studied the  $\chi_1^\pm\chi_2^0$ associated production,
 with the consequent decays of $\chi_1^\pm\rightarrow   \nu_\ell \ell \chi_1^0$ and $\chi_2^0
 \rightarrow \ell \ell \chi_1^0$ via on-shell sleptons.
 The invariant mass
of the lepton pairs, $m_{\ell\ell}$, from the neutralino decay has a
distinctive triangle shape with a sharp kinematic cutoff.  We discuss the
utilization of this triangle shape in $m_{\ell\ell}$ distribution to
identify the slepton signal.  We studied the trilepton plus missing $E_T$ signal and
obtained the effective cross section, $\sigma \times {\rm BR} \times {\rm acceptance}$, that is
needed for a 5$\sigma$ discovery as a function of the cutoff mass for
the LHC with center of mass energy 14 TeV and 100 ${\rm fb}^{-1}$
integrated luminosity.  Our results are model independent such that
they could be applied to other models with similar decay topology.
When applied to the MSSM   under simple assumptions, it is found that with 100 ${\rm
  fb}^{-1}$ integrated luminosity, a discovery reach in the left-handed slepton mass of about 600 GeV could be reached, which extends far beyond the slepton mass
reach in the usual Drell-Yan studies.

\end{abstract}


\maketitle

\section{Introduction}
\label{sec:intro}

While the Large Hadron Collider (LHC) has great potential in searching for
strongly interacting particles, its reach in the electroweak sector of
new physics scenarios is limited due to the suppressed electroweak
production cross sections.  Although those electroweak particles could
appear in the cascade decay of heavier colored objects, the discovery
reach depends strongly on the mass scale of the colored ones.  Current
LHC searches in the Minimal Supersymmetric Standard Model (MSSM) are mostly
focused on the direct pair production of squarks and gluinos.  Null
search results from both   ATLAS and CMS \cite{ATLAS, ATLAS2L,CMS, CMSOS2L, CMSmultilep} already
exclude the mass of those colored particles up to about 800 GeV.  It
is likely that those colored particles are so heavy that they are out
of the reach of the LHC.  In this paper, we consider the direct
production of the electroweak sector of the MSSM.  In particular, we
focus on the LHC reach for the discovery of the sleptons.  A complementary study on
the LHC reach in the neutralino and chargino sector with decoupled sleptons can be found in
Ref.~\cite{padhi}.

If low energy supersymmetry is realized in the nature, sleptons are
likely to be light.  This happens in the Gauge Mediated Supersymmetry (SUSY) breaking
scenarios \cite{GMSB}, as well as the Anomaly Mediated SUSY breaking
scenarios \cite{AMSB}, in which the slepton masses are proportional to
the electroweak gauge couplings.  Even in the minimal Gravity Mediated SUSY breaking scenarios (mSUGRA)
\cite{mSUGRA} where all the scalars have a common mass $m_0$ at some high
energy scale, renormalization group running to low energies typically
pushes up the squark mass (due to the contributions of strongly
interacting gluinos) while the sleptons remain light.  While the
masses of the superpartners of the colored objects have already been
constrained by current search limits, it is timely to fully explore
the discovery potential of the LHC for the superpartners of leptons.

In the $R$-parity conserving MSSM with the lightest neutralino $\chi_1^0$ being the
lightest supersymmetric particle (LSP), $\chi_1^0$ is a good candidate for
Weakly Interacting Massive Particle (WIMP) dark matter \cite{neutralinoDM}.  When sleptons are light, the
$t$-channel diagram mediated by the exchange of sleptons is important
in determining the annihilation cross section of $\chi_1^0$ dark
matter \cite{sleptonrelic}.  Therefore, discovery of the sleptons is
not only a verification of low energy supersymmetry in nature; precise
measurement of their masses also plays an important role in
determining the relic density of the neutralino LSP.

Earlier studies on the slepton discovery potential at the LHC mostly
focused on the   Drell-Yan pair production of slepton pairs, with
sleptons directly decaying down to lepton and $\chi_1^0$ \cite{earlyslepton, Baer,
  Bityukov}.  Most of those studies are   done either in the mSUGRA
framework or for a certain set of benchmark points only.  The
Drell-Yan production cross sections for slepton pairs are typically
small, suppressed both by the electroweak coupling strength, as well
as the scalar nature of the sleptons.  The LHC reach is very limited:
$m_{\tilde\ell_L} \gtrsim 300$ GeV and $m_{\tilde\ell_R} \gtrsim 200$
GeV for the LHC with center of mass 14 TeV and 30 ${\rm fb}^{-1}$
integrated luminosity \cite{Bityukov}.

In our study, we focused on an alternative production mechanism for
the sleptons via the on-shell decay of heavier neutralino and chargino states \cite{Baergaugino}. In particular,  we considered the scenario where $M_1< m_{\tilde{\ell}_L} < M_2 \ll  \mu$ and
studied the pair production of  Wino-like $\chi_1^\pm \chi_2^0 $ with the subsequent
decay of $\chi_2^0 \rightarrow \ell \tilde\ell_L \rightarrow
\ell\ell\chi_1^0$ and $\chi_1^\pm \rightarrow \ell \tilde\nu, \nu
\tilde\ell_L \rightarrow \ell \nu \chi_1^0$.  The collider signature
is trilepton plus missing $E_T$.

Compared to the traditional  
searches of slepton Drell-Yan pair production  with dilepton plus
missing $E_T$ signature, this channel is advantageous for the
following reasons.  Firstly, the production cross section for
$\chi_1^\pm \chi_2^0 $, although also at the electroweak strength, is
larger than slepton pair production due to the fermionic
nature of the neutralinos and charginos.  Secondly, for $\chi_2^0$ and
$\chi_1^\pm$ being dominantly Wino, the decay branching fraction into
left-handed sleptons is almost 100\%.
Thirdly, the SM backgrounds for this trilepton signature are also much smaller, with dominant
contributions from the leptonic decay of $WZ/\gamma^*$, as well as $t\bar{t}$
with an extra lepton from heavy flavor decay.

In addition, the invariant mass of the dileptons from the
$\chi_2^0$ cascade decay chain, $m_{\ell\ell}$, has a distinctive
triangle shape, with a sharp kinematic cutoff, $m_{\rm cut}$ completely
determined by $m_{\chi_2^0}$, $m_{\chi_1^0}$ and $m_{\tilde{\ell}_L}$ \cite{Hinchliffe:1996iu}.  
This triangle feature has been mostly used
as a precise determination of the slepton mass \cite{Hinchliffe:1996iu, Birkedal:2005cm, Abdullin:2006jm}.   
It is also used in the
recent CMS opposite sign dilepton searches to enhance the signal acceptance \cite{CMSOS2L}.  
Although this spectral shape can  be obvious to the eye,  it could easily get
washed out when SM backgrounds are considered.  In our study,
we   explore   the LHC discovery potential for
sleptons by performing a fit to this triangle spectral shape.  We also
fit the dilepton invariant mass distribution from the dominant SM
backgrounds (either containing a $Z/\gamma^*$ or from $t\bar{t}$).  
Since we keep the overall normalization of the backgrounds that contain a $Z$ as a fitting
parameter, our treatment  allows us to include
other backgrounds that contain a $Z$ as well, for example, those from other SUSY
processes.  We obtain the effective  
cross section, $\sigma\times$ BR $\times$ acceptance,  necessary for a 5$\sigma$ discovery
as a function of the cutoff mass.  This result is model independent
and can be applied to any new physics model that gives rise to the
same decay topology and signature.  When applied to the MSSM slepton
produced in the cascade decay of Wino-like $\chi_1^\pm \chi_2^0$, we obtain the
5$\sigma$ reach in the parameter space of $M_2$
vs. $m_{\tilde\ell_L}$.

A recent analysis by ATLAS on the same sign dilepton  plus missing $E_T$ signature \cite{ATLAS2L}
studied $\chi_1^\pm\chi_2^0$ associated production with decays of $\chi_1^\pm$ and $\chi_2^0$ via on-shell slepton in a simplified weak gaugino production model.  With 1 ${\rm fb}^{-1}$ at the 7 TeV LHC,  masses of $\chi_1^\pm$ ($\chi_2^0$) up to 200 GeV 
for $m_{\chi_1^0}=0$ GeV were excluded at 95\% C.L., based on event counting.  Fitting the dilepton invariant mass distribution was also performed in the opposite sign dilepton plus missing $E_T$ search at   CMS \cite{CMSOS2L}.  With 0.98 
${\rm fb}^{-1}$ integrated luminosity at the 7 TeV LHC, 95\% C.L. upper limits on the cross section times acceptance 
of about 4 $-$ 30 fb is obtained for the cutoff mass scale between 20 to 300 GeV.

The outline of the paper is as follows.  In Sec.~\ref{sec:MSSM}, we
discuss the slepton production and decay, focusing on the sleptons
produced via neutralino/chargino cascade decay and identify their collider
signature.  In Sec.~\ref{sec:mll}, we present the triangle spectral
shape of the dilepton invariant mass distribution and discuss the
parameter dependence of the cutoff mass.  In Sec.~\ref{sec:bounds}, we
review the current collider bounds on sleptons as well as bounds on
neutralinos and charginos in the MSSM.  In Sec.~\ref{sec:fit}, we discuss
in detail our treatment of the signal, as well as the SM backgrounds
by fitting to the spectral shape of the $m_{\ell\ell}$ distribution.
In Sec.~\ref{sec:results}, we present the model-independent least effective cross section, 
$\sigma\times {\rm BR} \times {\rm acceptance}$, that is needed for a
5$\sigma$ discovery as a function of cutoff mass at the LHC with center of mass energy 14 TeV and
100 ${\rm fb}^{-1}$ luminosity.  In Sec.~\ref{sec:reach}, we apply
our study in the MSSM electroweak sector and show the 5$\sigma$ reach in $M_2$
vs. $m_{\tilde\ell_L}$ parameter space.  In Sec.~\ref{sec:conclusion},
we conclude.

\section{MSSM with light gauginos and sleptons}
\label{sec:MSSM}

We consider the low lying spectrum of the MSSM electroweak sector,
which includes only neutralinos, charginos, and  light sleptons.
In our discussion below, we assume the canonical case where $M_1 <
M_2,  \mu$ ($M_1$, $M_2$ and $\mu$ being the mass parameters for Bino, Winos and Higgsinos, respectively) and the lightest neutralino LSP is dominantly Bino-type.
Scenarios with other mass orderings can be studied similarly.
The slepton mass spectrum is determined by $({\bf M}_{\tilde\ell}^2)_{LL}$,
$({\bf M}^2_{\tilde\ell})_{RR}$ and $({\bf M}_{\tilde\ell}^2)_{LR}$, where each
${\bf M}_{\tilde\ell}^2$ is a $3\times 3$ matrix, representing three generations.  In
our study, we take the simple assumption that the flavor mixing
between generations is negligible. The phenomenology and implication of
sizable flavor mixing in the slepton sector can be found in
Ref.~\cite{sleptonflavor,Krasnikov:1996np}.  Furthermore, the left-right mixing in the
slepton sector is typically proportional to the lepton Yukawa, which
is small for the first two generations.  Therefore, we assume there is
no left-right mixing for selectrons and smuons and label the mass
eigenstates as $\tilde\ell_L$ and $\tilde\ell_R$, for $\ell=e,\ \mu$,
with masses $m_{\tilde\ell_L}$ and $m_{\tilde\ell_R}$, respectively.
For the stau,   left-right mixing could be sizable, especially
with large $\tan\beta$.  The mass eigenstates are   labeled as
$\tilde\tau_1$ and $\tilde\tau_2$.  There are three parameters
involved for the stau sector: $m_{\tilde\tau_1}$, $m_{\tilde\tau_2}$
and the left-right mixing angle $\theta_{\tilde\tau}$.  Our discussion
below applies to the simplified case of selectrons and smuons,
although it could be  adapted to the stau case as well.  For
sneutrinos, their masses $m_{\tilde\nu_\ell}$ are also determined by $({\bf
  M}_{\tilde\ell})_{LL}^2$.  Therefore, their masses are related to
$m_{\tilde\ell_L}$ with a small splitting introduced by
electroweak effects: $m_{\tilde\nu_\ell}^2 = m_{\tilde\ell_L}^2 +
m_W^2\cos 2 \beta $; for the allowed range of $\tan\beta > 1$,
$m_{\tilde\nu_\ell} < m_{\tilde\ell_L}$.

The direct production channels of sleptons are Drell-Yan pair
productions of $\tilde{\ell}_L\tilde\ell_L$,
$\tilde{\ell}_L\tilde\nu_\ell$, $\tilde{\nu}_\ell\tilde\nu_\ell$ and
$\tilde{\ell}_R\tilde\ell_R$.  The production cross sections are
typically small due to both the electroweak coupling strength and the
scalar nature of the particles.  At the LHC with $\sqrt{s}=14$ TeV,
the cross sections vary from 0.2 pb to 0.5 fb for $\tilde{\ell}_L\tilde\ell_L$
and $\tilde{\nu}_\ell\tilde\nu_\ell$, from 0.8 pb to 1.5 fb for
$\tilde{\ell}_L\tilde\nu_\ell$, and from 0.08 pb to 0.2 fb for
$\tilde{\ell}_R\tilde\ell_R$, for masses of sleptons in the range of 100 to 500 GeV \cite{Baer}.

The decay of right-handed sleptons is quite straightforward, proceeding dominantly
into $\ell \chi_1^0$.  In cases when on-shell decay of $\tilde\ell_R$
into higher neutralino states is open and when there is a significant Bino
component in higher neutralino states (typical for $\mu \sim
M_1$), $\tilde\ell_R \rightarrow \ell \chi_{2,3,4}^0$ could also
contribute, although the decay branching fraction is almost always
suppressed.

The decay of left-handed sleptons depends on the neutralino and chargino  spectrum,
in particular, that of Wino-type neutralino and charginos.  Since the
decay of sleptons into Higgsino-type neutralinos and charginos are
typically suppressed, we assume $M_2 \ll \mu$, and thus $\chi_2^0$,
$\chi_1^\pm$ are mostly Wino-like.  For
$m_{\tilde\ell_L},\ m_{\tilde\nu_\ell} < M_2$,  the branching fractions of $\tilde\ell_L
\rightarrow \ell \chi_1^0$, $\tilde\nu_\ell \rightarrow \nu_\ell
\chi_1^0$ are almost 100\%.  Once
$m_{\tilde\ell_L},\ m_{\tilde\nu_\ell} \gtrsim M_2$,  the decays of $\tilde\ell_L
\rightarrow \ell \chi_2^0,\  \nu_\ell \chi_1^\pm$, $\tilde\nu_\ell
\rightarrow \nu_\ell \chi_2^0, \ \ell \chi_1^\pm$ become dominant.  The
branching fraction is about 10\% into $\chi_1^0$, 30\% into
$\chi_2^0$, and 60\% into $\chi_1^\pm$.  With the subsequent decay of
$\chi_2^0 \rightarrow Z^{(*)} \chi_1^0, \ h \chi_1^0$ and $\chi_1^\pm \rightarrow
W^{(*)} \chi_1^0$, left-handed slepton and sneutrino decay would have
multi-lepton, multi-jets, plus missing $E_T$ final states.  
 
For the Drell-Yan pair production of sleptons with dominant direct
decay of sleptons into $\chi_1^0$, the collider
signatures are dilepton plus missing $E_T$  for $\tilde{\ell}_L\tilde\ell_L$
and $\tilde{\ell}_R\tilde\ell_R$, single lepton plus missing $E_T$ for
$\tilde{\ell}_L\tilde\nu_\ell$, and missing $E_T$ only for
$\tilde{\nu}_L\tilde\nu_L$.  The single lepton channel suffers from
large SM backgrounds, mainly $W$.  The missing $E_T$  
only signature from $\tilde{\nu}_L\tilde\nu_L$  needs an
extra jet or lepton from initial or final state radiation, which leads
to further suppression of signal cross sections.  Current collider
analyses of slepton Drell-Yan production focus on the final states of
two isolated energetic leptons plus missing $E_T$ \cite{Baer, Bityukov}.  The
SM backgrounds are typically large, dominantly from $WW$ or $t\bar{t}$.  The LHC reach is very
limited: $m_{\tilde\ell_L} \gtrsim 300$ GeV and $m_{\tilde\ell_R}
\gtrsim 200$ GeV for the LHC with center of mass 14 TeV and 30 ${\rm
  fb}^{-1}$ integrated luminosity \cite{Bityukov}.

\begin{figure}
\begin{center}
\includegraphics[width = 0.45\textwidth]{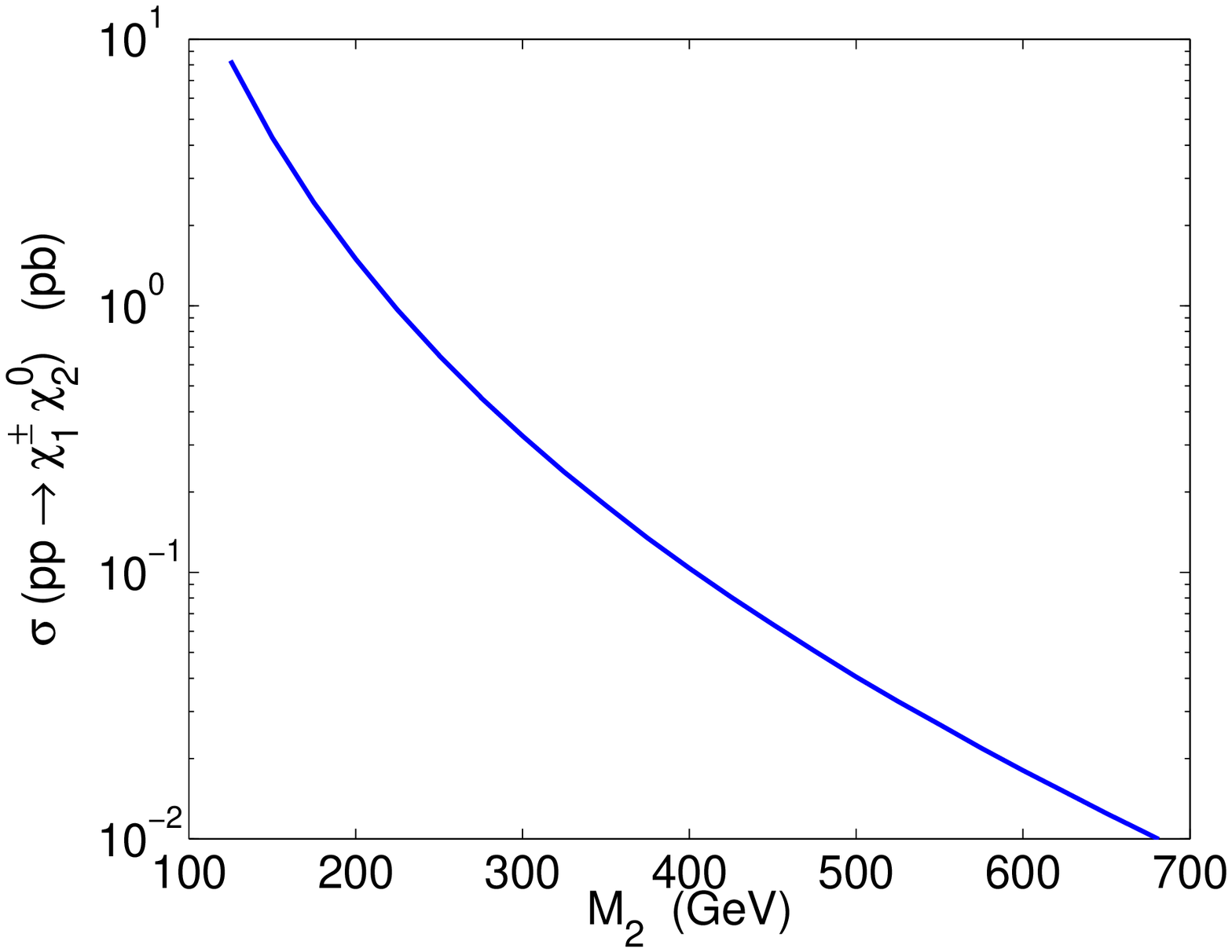}
\resizebox{3.15 in}{!}{\includegraphics*[0,550][320,750]{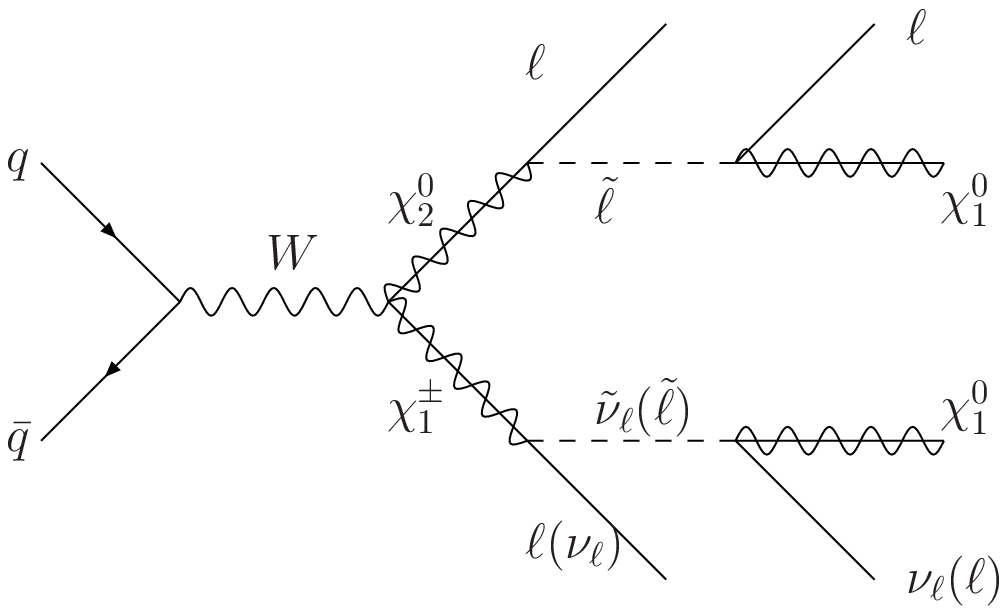}}
 \end{center}
\caption{The left plot shows the cross section for Wino-like 
$\chi_1^\pm \chi_2^0$ associated production at the LHC with center of mass energy 14 TeV.  
Here we have decoupled both Higgsinos as well as squarks.  
The right plot shows the Feynman diagram that gives rise to the trilepton plus missing
 $E_T$ final states.   }
\label{fig:Winoproduction}
\end{figure}

  In this paper, we explore alternative production channels for
sleptons, in particular,  $\tilde\ell_L$ and $\tilde\nu_\ell$, 
via   the decay of heavier neutralinos and charginos. The
coupling of Higgsinos to sleptons are highly suppressed by the small
lepton Yukawa coupling, therefore the production of sleptons from
Higgsino decay is negligible.  We assume $\mu$ is heavy and decouple
Higgsinos.  Winos, on the other hand, could dominantly decay into
$\tilde\ell_L$ and $\tilde\nu_\ell$ once it is kinematically available,
since the competing processes of $\chi_2^0 \rightarrow Z \chi_1^0$,
$ h \chi_1^0$ and $\chi_1^\pm \rightarrow W
\chi_1^0$ suffer from small neutralino mixing.    The pair production cross
sections of Wino-like neutralinos/charginos are larger compared to those of
sleptons of similar mass.  In Fig.~\ref{fig:Winoproduction}, we show
the production cross section for Wino-like $\chi_1^\pm \chi_2^0$ for the LHC
with $\sqrt{s}=14$ TeV.     The cross section
is about 10 pb for $M_2$ around 100 GeV, which drops to about 10 fb
for $M_2$ around 700 GeV.    In principle, $\tilde\ell_{L,R}$ could also appear in Bino-like neutralino
decay, in cases of $M_1> m_{\tilde\ell}, M_2,\mu$.  However, the pair productions of Bino-type neutralino with other neutralinos/charginos are typically suppressed due to the small neutralino mixing effects.

For $M_1 < m_{\tilde{\ell}_L}, m_{\tilde{\nu}_\ell} < M_2$ , the
lightest chargino $\chi_1^\pm$ dominantly decays into $\ell
\tilde\nu_\ell$ and $\nu_\ell \tilde\ell_L$.  With the consequent
decay of $\tilde\nu_\ell$ and $\tilde\ell_L$ directly into $\chi_1^0$,
the branching fraction of $\chi_1^\pm \rightarrow \ell \nu_\ell
\chi_1^0$ is almost 100\%.  $\chi_2^0$, on the other hand, decays into
$\nu_\ell \tilde\nu_\ell $ and $\ell \tilde\ell_L$ with about the same
branching fraction.  The former decay leads to $\nu \nu \chi_1^0$
final states, while the latter process has two isolated charged
leptons $\ell \ell \chi_1^0$.  Considering trilepton plus missing $E_T$
signatures, the overall branching fraction of $\chi_1^\pm \chi_2^0$
into this final state is about 50\%.  Combining the production cross
section of $\chi_1^\pm \chi_2^0$, left-handed sleptons could be
produced in the decay products of Wino-like heavier neutralino and
charginos states with relatively large cross sections compared to the
direct Drell-Yan process.   Relatively small SM background for the trilepton final state and the distinctive triangle spectral shape for $m_{\ell\ell}$ render this channel useful in probing the left-handed  sleptons  at the LHC.   Of course it
should always be kept in mind that such slepton production is only
possible when the slepton masses are less than $M_2$.

It should also be noted that such slepton production via cascade decay
of heavier neutralinos and charginos only  works effectively for the left-handed
charged sleptons.  For the right handed sleptons, even
if it is lighter than $M_2$, the branching fraction is small in
general compared to the dominant decays of $\chi_2^0 \rightarrow Z
\chi_1^0$, $\chi_2^0 \rightarrow h \chi_1^0$ since it is suppressed by
the small Wino-Bino mixing.  It only becomes important in the limited
parameter regions with small $M_2 - M_1$ such that the on-shell decay
of $\chi_2^0$ into the dominant channels are forbidden.  Therefore, in
our study below, we focus on the light left-handed sleptons.

In cases when $\mu$ is lighter: $M_1<\mu<M_2$, additional decay modes
of heavier Wino states into lighter Higgsino states plus Higgses open,
which could lead to the suppression of the branching fractions of
Winos decaying into sleptons.  The results that we obtained in our study,
however, can also be  applied to such cases, taking into
account the suppressed branching fractions.

\section{$m_{\ell\ell}$ distribution and triangle shape}
\label{sec:mll}

\begin{figure}
\begin{center}
\includegraphics[width = 0.5\textwidth]{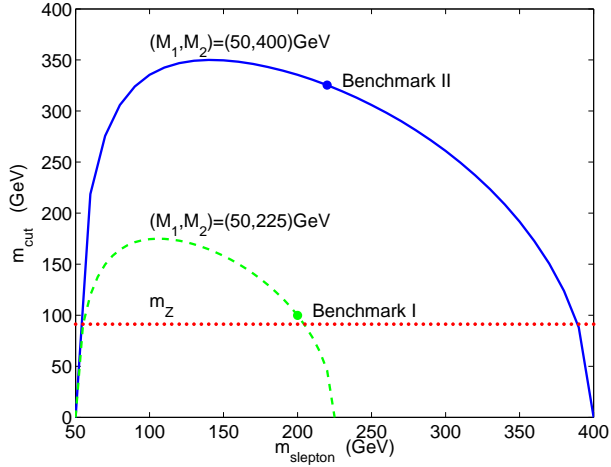}
 \end{center}
\caption{The dependence of the dilepton invariant mass $m_{\ell\ell}$ distribution endpoint $m_{\rm cut}$ 
on the  slepton mass $m_{\tilde{\ell}}$, for  $(M_1,M_2)$=(50,400) GeV 
(solid  curve)  and (50,225) GeV (dashed curve).
   Benchmark point I: $(M_1, M_2, m_{\tilde{\ell}_L})=(50, 225, 200)$ GeV    
   and benchmark point II: $(M_1, M_2,  m_{\tilde{\ell}_L})=(50, 400, 220)$ GeV are
     displayed as dots.  Also plotted is the $Z$ boson mass in red dotted line.  
}
\label{fig:mll}
\end{figure}

One distinctive feature of the dileptons from the on-shell cascade
decay of $\chi_2^0$ (see the right plot of
Fig.~\ref{fig:Winoproduction}) is that the invariant dilepton mass
$m_{\ell\ell}$ distribution has a triangle shape \cite{Hinchliffe:1996iu}, with a cutoff mass determined by the masses
of the $\chi_{1,2}^0$ and $\tilde{\ell}$:
\begin{equation}
m_{\rm cut}=m_{\chi_2^0}\sqrt{1-\frac{m_{\tilde{\ell}}^2}{m_{\chi_2^0}^2}}\sqrt{1-\frac{m_{\chi_1^0}^2}{m_{\tilde{\ell}}^2}}.
\label{eq:mcutoff}
\end{equation}
Fig.~\ref{fig:mll} shows the dependence of $m_{\rm cut}$ on
$m_{\tilde{\ell}}$ for a given set of $(M_1,M_2)$.  $m_{\rm cut}$
varies in the range of 0 to $M_2 - M_1$.  For $ 
M_2-M_1 > m_Z$, $m_{\rm cut}$ is larger than $m_Z$ (indicated by the straight red dotted line in
Fig.~\ref{fig:mll})  for a large range of $m_{\tilde{\ell}}$. 
 This is advantageous since we can effectively suppress the dominant SM
background from $WZ/\gamma^*$ by imposing a $m_{\ell\ell}$ cut to be above
$m_Z$.

This feature in the $m_{\ell\ell}$ spectral shape is often used as a
precise determination of the slepton mass
\cite{Hinchliffe:1996iu}.  Even
in the case of off-shell sleptons in neutralino decay, there have been
studies in the literature exploring the sensitivity of the
$m_{\ell\ell}$ spectral shape on the slepton mass
\cite{Birkedal:2005cm, Abdullin:2006jm}.  In our study below, we
explore how to use this distinctive triangle shape of the
$m_{\ell\ell}$ distribution to identify the slepton signal from the SM
backgrounds.  This triangle shape in $m_{\ell\ell}$ is not unique to
the specified $\chi_2^0$ decay in the MSSM;  it could appear in many
new physics model with a similar cascade decay topology   
that gives rise to two leptons.     Our analysis is therefore model
independent and can be   applied to a more general set of new
physics models.   
 
\section{Current collider search limits}
\label{sec:bounds}
The current best limits on the slepton masses come from LEP searches for  dilepton plus missing energy signatures \cite{LEPslepton}
with $\sqrt{s}$ up to 208 GeV.  For a mass splitting between slepton
and neutralino LSP above 15 GeV and considering only the contribution
from right-handed sleptons, the mass limits are: $m_{\tilde{e}} > 99.6$
GeV, $m_{\tilde{\mu}} > 94.9$ GeV and $m_{\tilde{\tau}} > 85.9$ GeV.
This is conservative,  since the production cross section for the
left-handed sleptons is higher.  
 For stau,  it is possible have a large left-right mixing, which could decrease the
production cross section for the lightest stau pair.  A lower limit of
$m_{\tilde{\tau}} > 85.0$ GeV can be obtained when the production
cross section for the lightest stau is minimized.  It should be noted
that the slepton mass limits are obtained with $\mu=-200$ GeV and
$\tan\beta=1.5$, a point at which the neutralino mass limit based on
the LEP neutralino and chargino searches is the weakest, and the
selectron cross section is relatively small.  The gaugino mass unification
relation $M_1=(5/3)\tan^2\theta_W M_2$ is assumed, which is relevant in
fixing the masses and field content of the neutralinos.  Slepton mass
limits would change for a non-unified mass relation between $M_1$ and
$M_2$, since neutralinos appear in both the slepton decay final
states, as well as participating in the $t$-channel diagram for
selectron production.  For selectrons, $\tilde{e}_L\tilde{e}_R$
production is also possible via $t$-channel neutralino exchange.  In
the case where the $\tilde{e}_R - \chi_1^0$ mass splitting is small and
the usual acoplanar dilepton search is insensitive, a single lepton plus
missing energy search yields a lower limit on $m_{\tilde{e}_R}$ of 73
GeV, independent of
$m_{\chi_1^0}$ \cite{LEPsinglelep}.  For sneutrinos, a mass limit of
45 GeV can be deduced from the invisible $Z$ decay width \cite{:2005ema}.  
An indirect mass limit on sneutrinos could be derived
from the direct search limits on the charged slepton masses.

Since we consider the production of sleptons from heavier neutralino  decay, we also
summarize the current status of the neutralino and chargino sector. 
Charginos $\chi_1^\pm$ can be pair produced at LEP via $s$-channel
exchange of $Z/\gamma^*$ or $t$-channel exchange of $\tilde\nu_e$,
with destructive interference.  It decays to $f\bar{f}^\prime
\chi_1^0$ via a virtual $W$ or sfermion, or dominantly to
$f\tilde{f}^\prime$ when two body decay is kinematically accessible.
In the case of heavy sfermions and a mass splitting
$m_{\chi_1^\pm}-m_{\chi_1^0}$ of at least a few GeV, a robust chargino mass lower limit of
103.5 GeV can be obtained for sneutrino masses larger than 300 GeV
\cite{LEPchargino1}, assuming the gaugino mass unification relation.
For the case of small mass splitting between the lightest chargino and
neutralino LSP, limits have been obtained for the degenerate gaugino
region ($M_1 \sim M_2$): $m_{\chi_1^\pm}>91.9$ GeV for large sneutrino
mass, as well as the ``deep Higgsino" region ($|\mu| \ll M_{1,2}$):
$m_{\chi_1^\pm}>92.4$ GeV \cite{LEPchargino2}.  For lower sfermion
masses, the limit is worse due to the reduced pair production cross
section, as well as the reduction of selection efficiency due to the
opening   of   two body decay channels.  In particular, there is a
so called ``corridor" region where $m_{\chi_1^\pm} - m_{\tilde{\nu}_\ell}$
is small and the lepton from $\chi_1^\pm \rightarrow \ell \tilde{\nu}_\ell$ is
so soft that it can escape detection.  Associated production of
$\chi_1^0\chi_2^0$ can be studied in cases where the
chargino search becomes ineffective.  Limits on chargino and
neutralino masses for the light sfermion case, therefore, depend on the
sfermion spectrum.

For the lightest neutralino LSP, there is no general mass limit from
LEP if the gaugino mass unification relation is not imposed.  Production
via $s$-channel exchange of $Z/\gamma^*$ could be absent for a
Bino-like neutralino, and $t$-channel production could be negligible
for heavy selectrons.  An indirect mass limit on the neutralino LSP can be
derived from chargino, slepton and Higgs boson searches, when gaugino
mass (and sfermion mass) unification relation is assumed.  A lower
mass limit of 47 GeV can be obtained at large $\tan\beta$
\cite{LEPLSP}, while a tighter limit of 50 GeV can be derived in the
mSUGRA scenario \cite{LEPLSPmSUGRA}.

Trilepton searches at Tevatron Run II \cite{CDFtrilepton, Abazov:2009zi} study the associated production of 
$\chi_1^\pm \chi_2^0$ with the subsequent decay of 
$\chi_1^\pm \rightarrow \ell \nu \chi_1^0$ and $\chi_2^0 \rightarrow \ell^+ \ell^-\chi_1^0$.   $\sigma \times {\rm BR}(\chi_1^\pm\chi_2^0 \rightarrow 3 \ell)$ is bounded to be less than about 0.13 $-$ 0.06 pb (0.5 $-$ 0.1 pb) from D\O~(CDF) searches    for chargino mass in the range of 100 $-$ 180 GeV.   For sufficient light sleptons, the leptonic decay branching fractions are large and a mass limit on the lightest chargino can be derived based on the null search results.   
A chargino mass limit of 138 GeV is obtained based on D\O~searches, when the leptonic branching fraction for three body decay is maximized, while no mass limit can be derived in the large $m_0$ 
case \cite{Abazov:2009zi}.   
A recent CDF analysis with 5.8 ${\rm fb}^{-1}$ data gives a mass limit on the chargino to be 168 GeV at 95\% C.L., for the mSUGRA benchmark point $m_0=60$ GeV, $\tan\beta=3$ and $A_0=0$ \cite{CDFtrilepton}.

Limits on $\sigma \times {\rm BR}(\geq 3 \ell)$ are also obtained based on the recent trilepton search   from CMS collaboration using 2.1 ${\rm fb}^{-1}$ data collected at the LHC with $\sqrt{s}=7$ TeV \cite{CMSmultilep}.   No jet veto is imposed and the dominant contribution to the trilepton signal is from gluino cascade decay.  No limit on the chargino mass can be derived based on the direct pair production of  $\chi_1^\pm \chi_2^0$. 

Recent analyses by ATLAS on the same sign dilepton  plus missing $E_T$ \cite{ATLAS2L}
studied $\chi_1^\pm\chi_2^0$ associated production with the consequent decay of $\chi_1^\pm$ and 
$\chi_2^0$ via an on-shell slepton.   Assuming 
$m_{\tilde{\ell}}=\frac{1}{2}(m_{\chi_1^0} + m_{\chi_1^\pm})$, $m_{\chi_1^\pm}=m_{\chi_2^0}$, 
masses of $\chi_1^\pm$ ($\chi_2^0$) up to 200 GeV 
for $m_{\chi_1^0}=0$ GeV are excluded at 95\% C.L with 1 ${\rm fb}^{-1}$ data collected at the 7 TeV LHC.  
For $m_{\chi_1^0}$ about 50 GeV, the limit on $m_{\chi_1^\pm}$, $m_{\chi_2^0}$ is weakened to be about 150 GeV.

CMS performed an analysis on the opposite sign dilepton plus missing $E_T$ final states by looking for the kinematic edge in the dilepton invariant mass distribution \cite{CMSOS2L}.  With 0.98 
${\rm fb}^{-1}$ integrated luminosity at the 7 TeV LHC, 95\% C.L. upper limits on the cross section times acceptance 
of about 4 $-$ 30 fb are obtained for the cutoff mass scale between 20 to 300 GeV, assuming the signal efficiency of the LM1 benchmark point: $m_0=60$ GeV, $m_{1/2}=250$ GeV, $\tan\beta=10$, $A_0=0$ and $\mu>0$.

\section{Method}
\label{sec:fit}

The dominant standard model backgrounds for the trilepton plus missing $E_T$ signal are $WZ/\gamma^*$
and $t\bar{t}$ with a fake lepton (dominantly from $b$ decay).   
Note that trileptons from heavy flavor bottom and charm decay 
(produced in association with $Z/\gamma^*$)  could also be a significant background  
 \cite{Sullivan:2008ki}.    As explained below, the overall 
 normalization of backgrounds with a $Z$ peak is a fitting parameter in our analysis; 
 backgrounds containing heavy flavor produced in association with $Z/\gamma^*$ 
 could be included as well.  
Therefore, we don't simulate such heavy flavor backgrounds in our analyses.

Many SUSY models can
generate a trilepton signal, for example, $\chi_1^\pm \chi_2^0$ with
$\chi_2^0 \rightarrow \chi_1^0 Z^{(*)}$ and $\chi_1^\pm \rightarrow
\chi_1^0 W^{(*)}$.  For large mass splitting of
$m_{\chi_2^0}-m_{\chi_1^0} > m_Z$, the dilepton invariant mass
distribution from an on-shell $Z$ looks like that of   Standard Model $WZ$.  Such SUSY
backgrounds, if they exist, are included in our fitting to
$m_{\ell\ell}$ from the $Z$ pole, since the overall normalization of the $Z$
contribution is a fitting parameter.  For small mass splittings
$m_{\chi_2^0}-m_{\chi_1^0} < m_Z$, $m_{\ell\ell}$ is peaked near
$m_{\chi_2^0}-m_{\chi_1^0}$.  For such a case, a dedicated analysis to distinguish such off-shell $Z$ contributions from the triangle spectral shape is necessary.

The first difficulty in reconstructing the trilepton event is the
combinatorial ambiguity arising from the presence of a third lepton
in the final state. To resolve this issue we use the standard
technique of   same-sign subtraction. We construct our invariant mass
histograms by including both of the opposite-sign pairs of leptons and then
subtract the histogram of same-sign dilepton invariant mass to have a
good approximation to the histogram of invariant masses of the correct
pair of opposite-sign leptons, which is otherwise not experimentally
accessible.  While other techniques exist to resolve this ambiguity \cite{Rajaraman:2010hy},
they all sacrifice statistics for purity, and as our intent is to find
the general shape,  the statistics will generally be of more value for
us than the purity.

We model the dilepton invariant mass distribution by:
\begin{equation}
\frac{d\sigma}{dm_{\ell \ell}} = (f_{\rm triangle} * g) + f_Z + a_0f_{t\bar{t}}.
\label{eq:fitformulae}
\end{equation}
The first term models the triangle shape of the $m_{\ell \ell}$ from the signal process: 
\begin{equation}
f_{\rm triangle}(m_{\ell \ell}) = 
\left\{
\begin{array}{ll}
2N_{\rm sig}\frac{m_{\ell \ell}}{m_{\rm cut}^2}\ \ \ \ \ &0<m_{\ell\ell}<m_{\rm cut}\\
0& {\rm otherwise}
\end{array}
\right.
,
\end{equation}
where $N_{\rm sig}$ and $m_{\rm cut}$ are fit parameters for the
number of counts in the triangle and the mass of the cutoff.  To take
into account the detector effects, we smear the triangle by a
convolution of the triangle function with a gaussian with variance
$\sigma^2$ shown below:  
\begin{equation}
 (f_{\rm triangle} * g) \equiv \frac{1}{\sqrt{2\pi} \sigma} \int_{-\infty}^{\infty}dt\, e^{-\frac{t^2}{2\sigma^2}} f_{\rm triangle}(m_{\ell\ell}-t).
\end{equation}
 
The second term models the contribution to $m_{\ell \ell}$ from
processes that involves a $Z$ peak using the Breit-Wigner function:
\begin{equation}
f_Z(m_{\ell \ell}) = \frac{A}{2\pi} \frac{\Gamma }{(m_{\ell \ell}-m_0)^2+(\Gamma/2)^2}, 
\label{eq:zpole}
\end{equation}
where  $A,m_0,\Gamma$ are fit parameters for the amplitude,  centroid, and width of the $Z$ pole. 
 Note that we fit 
the centroid and the width of the $Z$ pole instead of using the SM values
 to account for the smearing effects introduced by the detector resolution.

The last term models the contribution to $m_{\ell \ell}$ from $t\bar{t}$ trilepton events, where 
$f_{t\bar{t}}$  is the $m_{\ell \ell}$ distribution from $t\bar{t}$ dilepton
events taken from Monte Carlo,  scaled by a factor $a_0$, a parameter 
to represent the fake rate of $t\bar{t}$ to  give three leptons. This is intended 
to emulate a data-driven background understanding, where the dilepton mass 
distribution is measured in a control region and then used to understand 
the background in the signal region. We choose the dilepton distribution 
because we expect that the two ``wrong" pairs will largely cancel each other 
out in the subtracted distribution.

Thus, we have in total seven fitting parameters:

\begin{itemize}
\item Number of events in triangle: $N_{\rm sig}$
\item Cutoff in triangle distribution: $m_{\rm cut}$
\item Amount of gaussian smearing of the triangle: $\sigma$
\item Amplitude of $Z$ peak: $A$
\item Apparent width of $Z$ peak: $\Gamma$
\item Apparent centroid of $Z$ peak: $m_0$
\item Fake rate for $t\bar{t}$ events: $a_0$
\end{itemize}

We use Madgraph 5 version v0.6.2 \cite{Alwall:2011uj} and Madevent v4.4.57 \cite{MadGraph} to generate our signal and background events.  These
events are passed to Pythia v6.4 \cite{PYTHIA}  to simulate initial state radiation, final state radiation, showering and
hadronization.  Additionally we use PGS4 \cite{PGS} with the ATLAS detector card to simulate detector
effects.  Note that, in producing and fitting the dilepton invariant
mass, we require only that there be three leptons in the event. In
particular, there is no requirement of low hadronic activity, which
means that strong production of particles which later decay through an
on-shell slepton can also be measured using this technique. We also do
not require any missing energy, which allows this technique to be
applicable in scenarios which do not include an invisible final state
particle, such as $R$-parity violating theories. 
In general,
some minimal requirement will be needed to ensure that events can be
triggered on (either through a single lepton trigger or dilepton trigger), but all of the points we consider have spectra which are not compressed enough to have significant loss due to triggering efficiencies.

For signal generation, we considered the simplified case where the
lightest neutralino is purely Bino in nature, and the second
neutralino is purely Wino, with degenerate charginos which are also
purely Wino: $m_{\chi_1^0}=M_1$ and  $m_{\chi_1^\pm}=m_{\chi_2^0}=M_2$.
This corresponds to taking the Higgsino mass parameter
$\mu$ to be heavy such that the Higgsinos decouple.  
We assume there is no left-right mixing, as well as no
flavor mixing between slepton generations.    We also completely decouple the heavy colored objects. 
 The relevant mass
parameters involved are $M_1$, $M_2$ and $m_{\tilde{\ell}_L}$.

We simulate the associated production of 
$\chi_1^\pm \chi_2^0$ with the consequent decay of $\chi_1^\pm$ and $\chi_2^0$
 via left-handed sleptons, as shown in Fig.~\ref{fig:Winoproduction}.  
 As discussed earlier in Sec.~\ref{sec:MSSM},  since $\chi_2^0$ has equal probability
 to decay into $\tilde\ell_L$ or $\tilde\nu_\ell$,  
 we obtain trilepton final states $\ell\ell\ell+\met$ 50\% of the time.   
 For simplicity, we only consider trilepton events with $\ell$ being either an electron or muon. 
In cases with lepton universality,
$m_{\tilde{e}_L}=m_{\tilde{\mu}_L}=m_{\tilde{\tau}_L}$, we could study
the same flavor, opposite sign $m_{\ell \ell}$ distribution.   For non-degenerate slepton masses between
generations, multiple triangle shapes appear and the analysis is more
complicated,  though with no flavor mixing they appear in different channels. 
Note that in the realistic case when the neutralino and chargino mass eigenstates
 are not the pure gauginos, the corresponding branching fractions for the decay
 into $\tilde\ell_L$ and $\tilde\nu_\ell$ need to be considered. 
For the backgrounds, we generate the SM $WZ/\gamma^*$ trilepton events, as well
as $t\bar{t}$ with both tops decaying  leptonically.

We construct our signal and background histograms of the dilepton
invariant mass by using our Monte Carlo data as probability distributions
from which we select events.  For the signal, we draw from the
opposite-sign lepton events 2$N_{\rm sig}$ times and draw from the
same-sign lepton events $N_{\rm sig}$ times and construct the
difference between the opposite-sign and same-sign distributions.
$N_{\rm sig}$ is the number of events that we are including in the
pseudoexperiment we are currently generating.  For a given luminosity,
$N_{\rm sig} = {\cal L} \times \sigma_{\rm sig} \times {\rm BR}\times {\rm \ acceptance}$.
Similarly, we can draw $N_{WZ/\gamma^*}$ events from the SM $WZ/\gamma^*$ background and
construct the corresponding $m_{\ell \ell}$ distribution.

Purely leptonic $t\bar{t}$ decay could lead to trilepton events with
a third faked lepton from $b$ decay.  In the pseudoexperiment we are
currently generating, we estimate the expected trilepton events
$N_{t\bar{t}}$ using the fake rate estimated from PGS simulation: $N_{
  t\bar{t}} = {\cal L} \times \sigma_{t\bar{t}} \times {\rm BR} \times {\rm fake\ rate}\times {\rm \ acceptance}$, where the fake rate is about $4.3\times 10^{-3}$.

In a trilepton $t\bar{t}$ sample, the opposite sign dilepton $m_{\ell
  \ell}$ distribution originating from $W^+W^-$ decay is the same as
  the dilepton distribution from dilepton $t\bar{t}$ events.  While
opposite-sign/same-sign dilepton $m_{\ell \ell}$ distributions with one
lepton from $W$ and the other from near/far $b$ jets have a very
different distribution in principle,  due to poor Monte-Carlo
statistics on the three lepton $t\bar{t}$ events, we draw from the opposite-sign dilepton  distribution of the 
 dilepton $t\bar{t}$ event samples $N_{t\bar{t}}$ times to simulate
the expected final distribution of $m_{\ell \ell}$ from $W$ pair
decay.  We take the simplified assumption that the same sign and
opposite sign $m_{\ell \ell}$ distributions from the $Wb$ combination are
very similar and build two random histograms from the 
same-sign dilepton distribution of the trilepton $t\bar{t}$ sample with $N_{t\bar{t}}$ entries each to simulate
the wrong pair of opposite-sign leptons and the same-sign pair of
leptons, respectively.  We then combine these histograms appropriately
to get our pseudoexperiment result histogram for the $t\bar{t}$
background. While this is not a perfect description of the background
in question due to physical differences between the distributions of
the incorrect pair of opposite-sign leptons and the pair of same-sign
leptons, it is approximately valid and a more accurate data-driven
background method than the one we use will be able to include those
differences without great difficulty, so the sensitivities we find
using this technique should be valid.

\begin{figure}
\begin{center}
\includegraphics[width = 3 in]{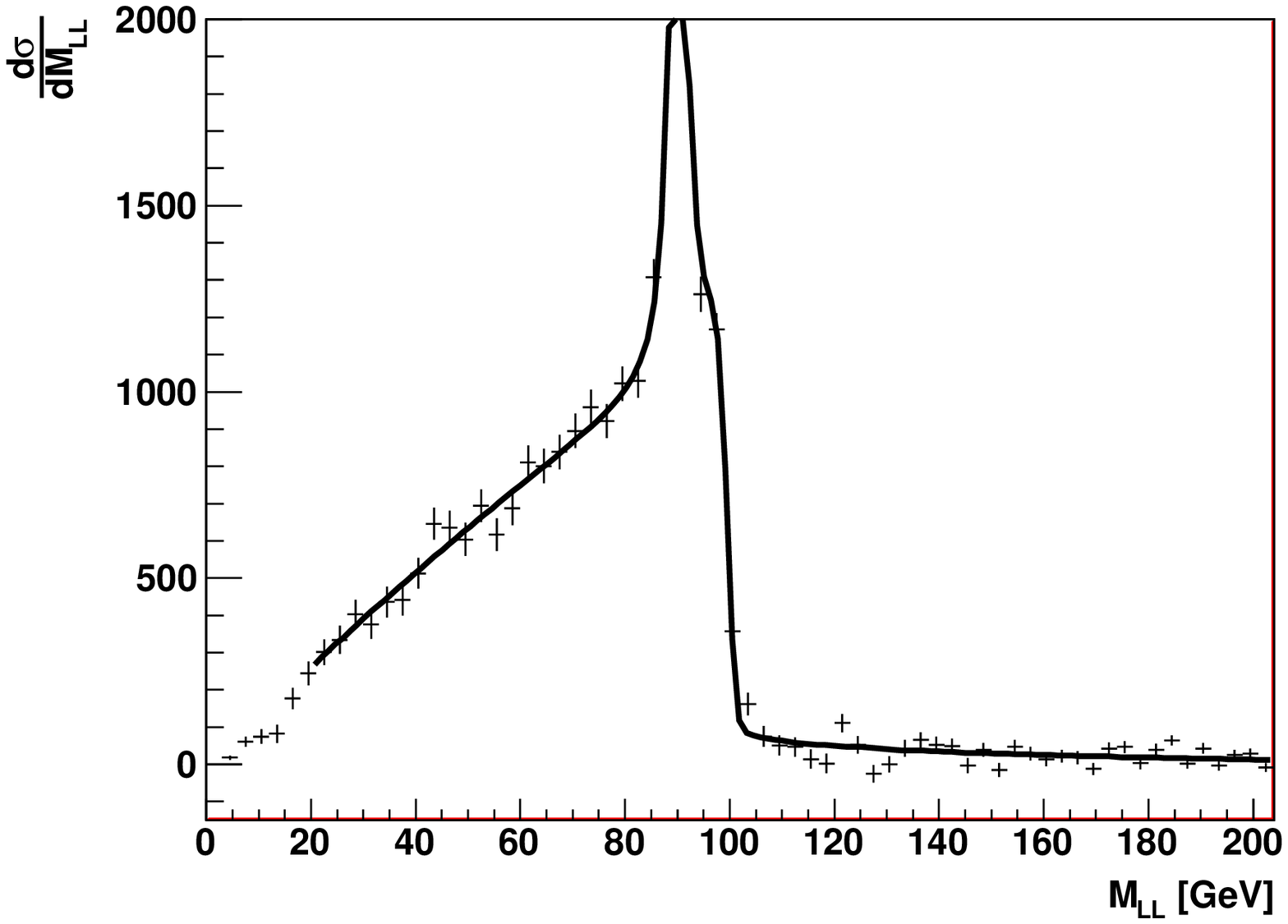}
\includegraphics[width = 3 in]{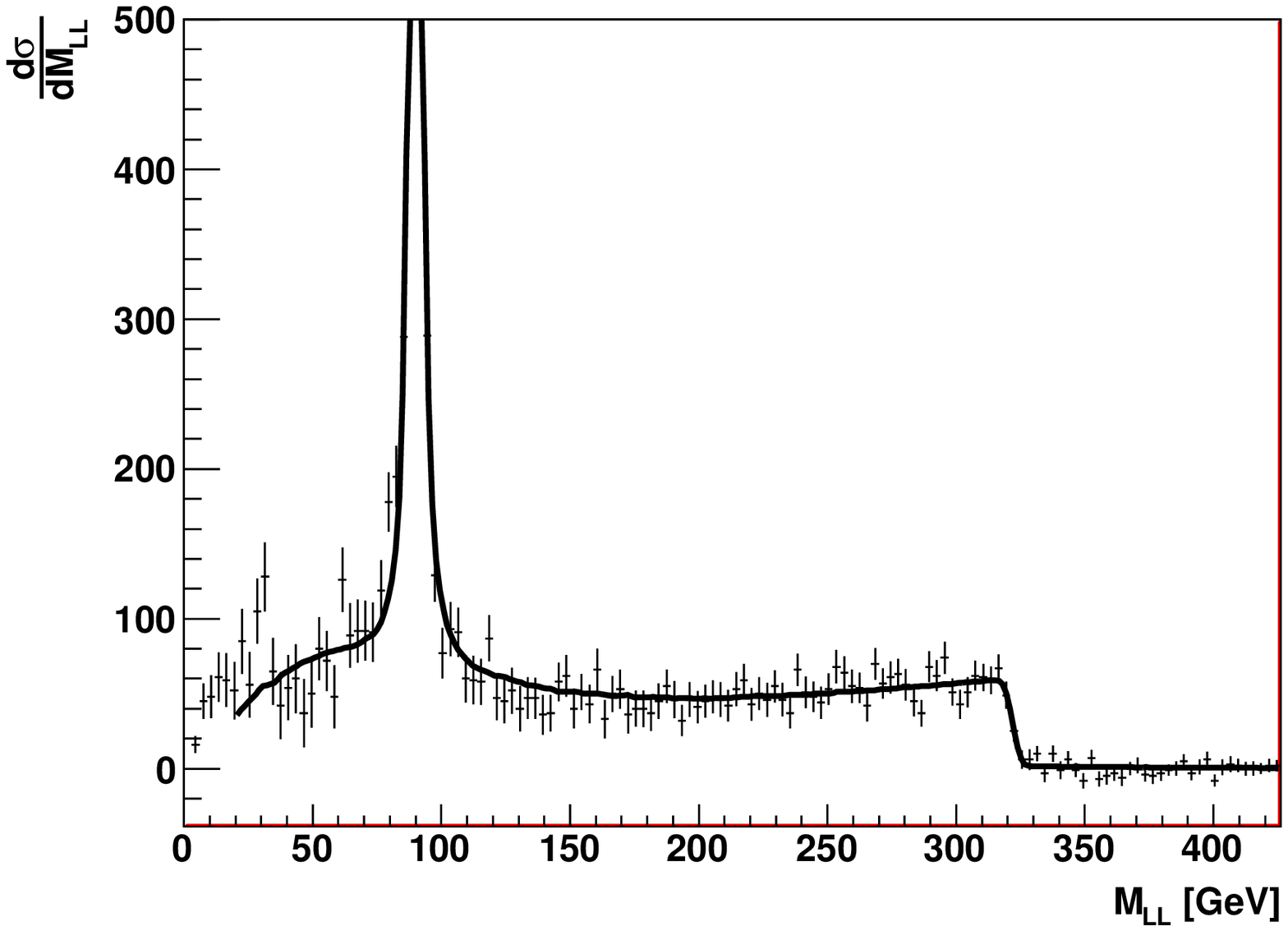}
\end{center}
\caption{$m_{\ell \ell}$ distribution for MSSM trilepton signal and
  dominant SM backgrounds at the 14 TeV LHC with an integrated luminosity
  of 100 ${\rm fb}^{-1}$.  The relevant  MSSM parameters are chosen to be
  benchmark point I: $(M_1, M_2, m_{\tilde{\ell}_L})=(50, 225, 200)$ GeV with $m_{\rm cut}=$100 GeV (left plot), 
  and benchmark point II: $(M_1, M_2,
  m_{\tilde{\ell}_L})=(50, 400, 220)$ GeV with $m_{\rm cut}=$325 GeV (right plot).  
    Also shown as black curves are the best
  fit distribution using Eq.~(\ref{eq:fitformulae}).   }
    \label{fig:fit}
\end{figure}

In Fig.~\ref{fig:fit}, we show the $m_{\ell \ell}$ distribution for the
MSSM trilepton signal and dominant SM backgrounds at the 14 TeV LHC
with integrated luminosity of 100 ${\rm fb}^{-1}$, for two benchmark
points: I, $(M_1, M_2, m_{\tilde{\ell}_L})=(50, 225, 200)$ GeV (left
plot); II, $(M_1, M_2, m_{\tilde{\ell}_L})=(50, 400, 220)$ GeV (right
plot).  The corresponding triangle cutoff masses are 100 GeV and 325 GeV,
respectively.   For $m_{\rm cut}$ near  
$m_Z$ (as in Benchmark Point I), the triangle distribution of $m_{\ell\ell}$ from the signal
process is buried under the SM $Z$ pole, 
making the identification of such triangle features much more difficult.  
For $m_{\rm cut}$ far above  $m_Z$ (as in Benchmark Point II), 
the sharp cutoff feature in $m_{\ell\ell}$ distribution can be easily 
identified from the SM background, as shown in the right plot of Fig.~\ref{fig:fit}.

For a given value of integrated luminosity, we fit the $m_{\ell \ell}$
distributions built from the signal and background events generated as
described above, using formulae given in Eqs.~(\ref{eq:fitformulae}) $-$ (\ref{eq:zpole})
with seven fitting parameters.  When the cutoff falls near the $Z$ pole,
we find that the fitting routine has too much freedom and often fails
to identify the cutoff in the proper location. This failure is due to
the similarity of the sharp feature of the cutoff and the edge of the $Z$
pole.  This results in a large degeneracy in the signal and $Z$ pole fit
parameters.  Therefore, we fix the $Z$ width, $\Gamma$ to be 3.121 GeV
and centroid position, $m_0$ to be 90.158 GeV when the cutoff is less than
150 GeV.  These values are obtained from the $m_{\ell \ell}$
distribution of the SM $WZ/\gamma^*$ backgrounds for events passed through the
PGS detector simulation.

We perform a $\chi^2$ fit
using the MINUIT fitting routine \cite{MINUIT}. 
We fit from 20 GeV to  ($m_{\rm cut} + $100 GeV), 
with a fixed binning scheme of 3 GeV/bin.  
  We find a good fit to the data with $\chi^2/{\rm dof} \sim
1$ for all cutoff masses.  For high cutoff masses, $\chi^2/{\rm dof}$ is
slightly lower due to the fact that   our fit spans more
bins. Conversely, we find a slightly higher $\chi^2/{\rm dof}$ at lower
cutoff mass. The data driven approach to fitting the background works
very well providing a best fit $\chi^2/{\rm dof}$ of about 1 when fitting
only the background.  For high cutoff masses, we also consider fits
from 125 GeV to ($m_{\rm cut}+$100 GeV).   This to ensure that
our fits for the cutoff at high mass are not being influenced too
greatly by fitting the $Z$ at low mass.  As an illustration, we show in
Table~\ref{table:fit} the input parameters for the simulation, as
well as the fitting parameters for the two benchmark points. 
  
  \begin{table}
\begin{tabular}{|c|c|c|c|c|c|c|c|c|c|} \hline
&&$A $&$\Gamma$ &$m_0$ &$a_0  $&$m_{\rm cut}$ &$N_{\rm sig}$&$\sigma$ &$\chi^2/{\rm dof}$ \\ 
&& &  (GeV)&(GeV)&$ (\times 10^{-3})$& (GeV)& & (GeV)&  \\ \hline 
 Benchmark &input&$3.981\times 10^3 $&3.121&90.158&4.280&99.8&$1.85\times 10^4$& &  \\ 
Point I &fitted&$3.299\times 10^3 $&(3.121)&(90.158)&5.420&99.5&$1.83\times 10^4 $&0.980&1.54 \\  \hline
 Benchmark&input&$3.981\times 10^3 $&3.121&90.158&4.280&325.3&$3.17\times 10^3 $& &   \\ 
Point  II &fitted&$3.347\times 10^3  $&2.914&90.041&4.412&321.9&$3.14\times 10^3 $&2.319&0.94  \\ \hline
\end{tabular}

\caption{Input and final fitting parameters to
  Eq.~(\ref{eq:fitformulae}) for two benchmark points: I, $(M_1, M_2,
  m_{\tilde{\ell}_L})=(50, 225, 200)$ GeV; II, $(M_1, M_2,
  m_{\tilde{\ell}_L})=(50, 400, 220)$ GeV, for the 14 TeV LHC with ${\cal L}=100\ {\rm fb^{-1}}$.   The cutoff for
  benchmark I is less than 150 GeV so we fix the width and centroid of
  the $Z$ peak with the values  obtained from the $m_{\ell \ell}$
distribution of the SM $WZ/\gamma^*$ backgrounds based on Monte-Carlo simulation. }
\label{table:fit}
\end{table}

\section{Results}
\label{sec:results}

Using the fitting strategy described above and assuming only the SM
backgrounds for trilepton plus missing $E_T$  signals, for a given cutoff mass $m_{\rm
  cut}$, we fit for the cutoff feature, marginalizing over the other
fit parameters.  This allows us to determine the number of events
required for a detection of the cutoff feature with 5$\sigma$-level
confidence.  We consider cutoff masses ranging from 75 to 650 GeV
stepping by 25 GeV, with finer stepping around the $Z$ pole.  As
explained before, for cutoff masses below 150 GeV, we perform a
five-parameter fitting with the $Z$ width and centroid fixed.  For
cutoff masses above 150 GeV, we perform a seven-parameter fitting,
allowing the fitting parameters of the $Z$ pole to vary freely.

\begin{figure}
\begin{center}
 \includegraphics[width = 0.45\textwidth]{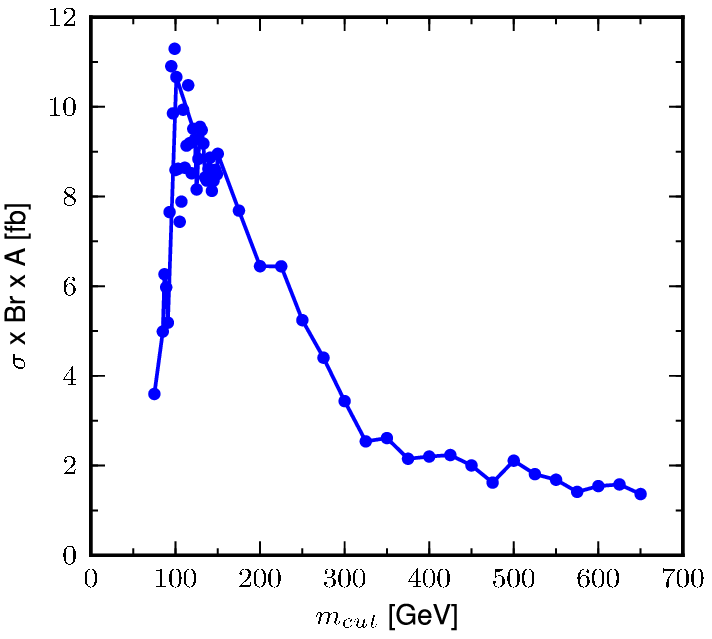}
\includegraphics[width = 0.45\textwidth]{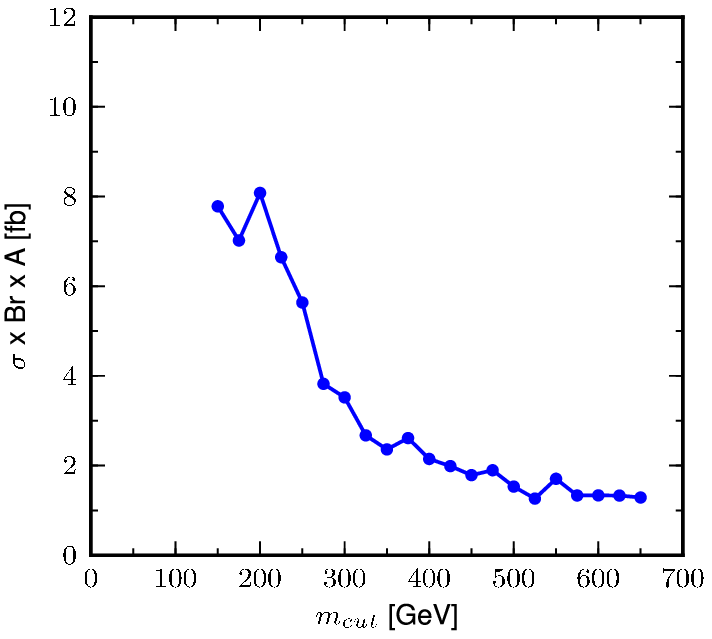}
\end{center}
\caption{ Effective trilepton cross section, $\sigma \times {\rm BR} \times {\rm acceptance}$,  at the 14 TeV LHC required
  for a 5$\sigma$ detection of the cutoff feature with an integrated luminosity of 100 ${\rm fb}^{-1}$.  
  For the left plot, the
  cross section reach is obtained with a fitting for $m_{\ell\ell}$ in
  the range of 20 GeV to ($m_{\rm cut} + $100 GeV).   
 For the right plot, a fitting range of 125 GeV to ($m_{\rm cut} + $100 GeV) is used.    }
 \label{fig:sigma_reach}
\end{figure}

In the left plot of Fig.~\ref{fig:sigma_reach} we plot the
required effective trilepton cross section, $\sigma \times {\rm BR} \times {\rm acceptance}$,  
as a function of cutoff mass for 5$\sigma$ discovery at
the 14 TeV LHC with an  integrated luminosity of 100 ${\rm fb}^{-1}$.  The maximum near 100 GeV is
due to the difficulty of detecting the cutoff feature near the $Z$ pole.
There is a significant parameter degeneracy between the amplitude of
the $Z$ and the number of counts in the triangle. The effect also
creates a significant scatter amongst the data points.
 At cutoff masses greater than 150 GeV, the required cross section for
detection decreases with increasing cutoff mass.  The scatter at high
cutoff mass is due to statistical fluctuations in the Monte Carlo
during the fitting process.
 The quality of the fit, however, depends on the range of $m_{\ell\ell}$
that is used in the fitting.  

In the right plot of Fig.~\ref{fig:sigma_reach}, we show the
5$\sigma$ reach in the effective cross section similar to the left plot, but
only fitting the $m_{\ell\ell}$ distribution above 125 GeV rather than
above 20 GeV, such that the background contributions are significantly
less relevant to the fit.  Comparing to the left plot, in which a
blind fitting is performed without prior knowledge of possible range
of $m_{\rm cut}$,  although the resulting lower limit on the effective
 cross section is very similar in both cases,
the fit with $m_{\ell\ell}>$ 125 GeV above the $Z$ pole is more robust
 since it greatly reduces the dependence on the precise knowledge of the backgrounds,
 in particular, those containing a $Z$ pole.

Note that the above result is obtained for an integrated luminosity of
100 ${\rm fb}^{-1}$.  There is no simple scaling behavior of the
required effective cross section for a different luminosity because
this fitting technique does not have the simple statistical behavior
 of a counting experiment. In order to understand the
sensitivity of this technique at significantly different luminosities
it is necessary to generate a new set of pseudoexperiments and analyze
them as explained above.

\section{Slepton Reach in MSSM}
\label{sec:reach}

The most straightforward application of this search is
to a system which gives the maximal branching ratio for the slepton
decay of the second neutralino.  We therefore consider the simplified
MSSM scenario that is described earlier, in
which $\chi_1^0$ is purely Bino and $\chi_1^\pm$, $\chi_2^0$ are
purely Winos with no left-right mixing in the slepton sector and squarks decoupled.   Note that in this simplified
scenario, such a trilepton search channel is only sensitive to the
intermediate $\tilde{\ell}_L$ since decays to $\tilde{\ell}_R$ are forbidden
due to the absence of couplings.

\begin{figure}
\begin{center}
 \includegraphics[width = 0.45\textwidth]{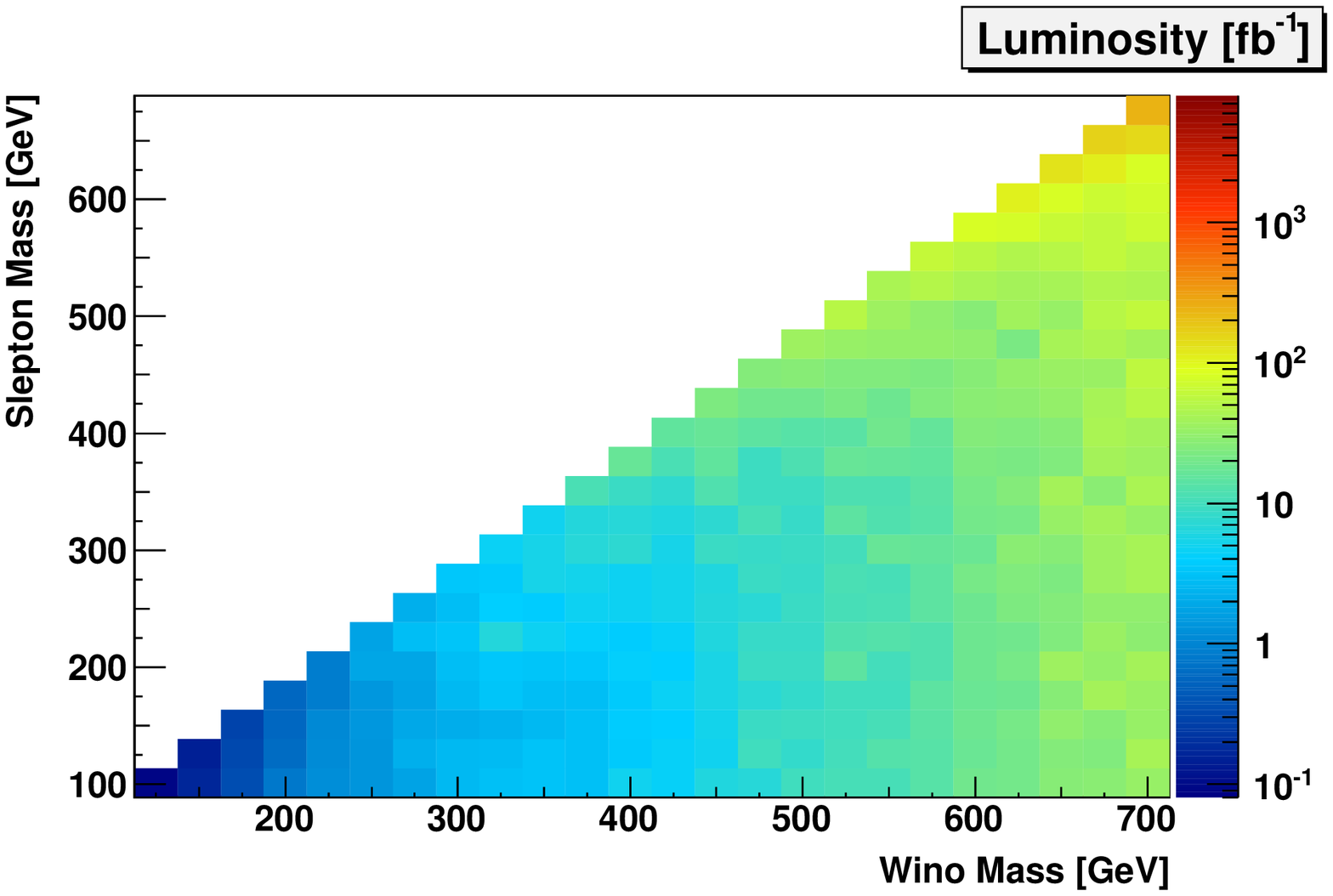}
\includegraphics[width = 0.45\textwidth]{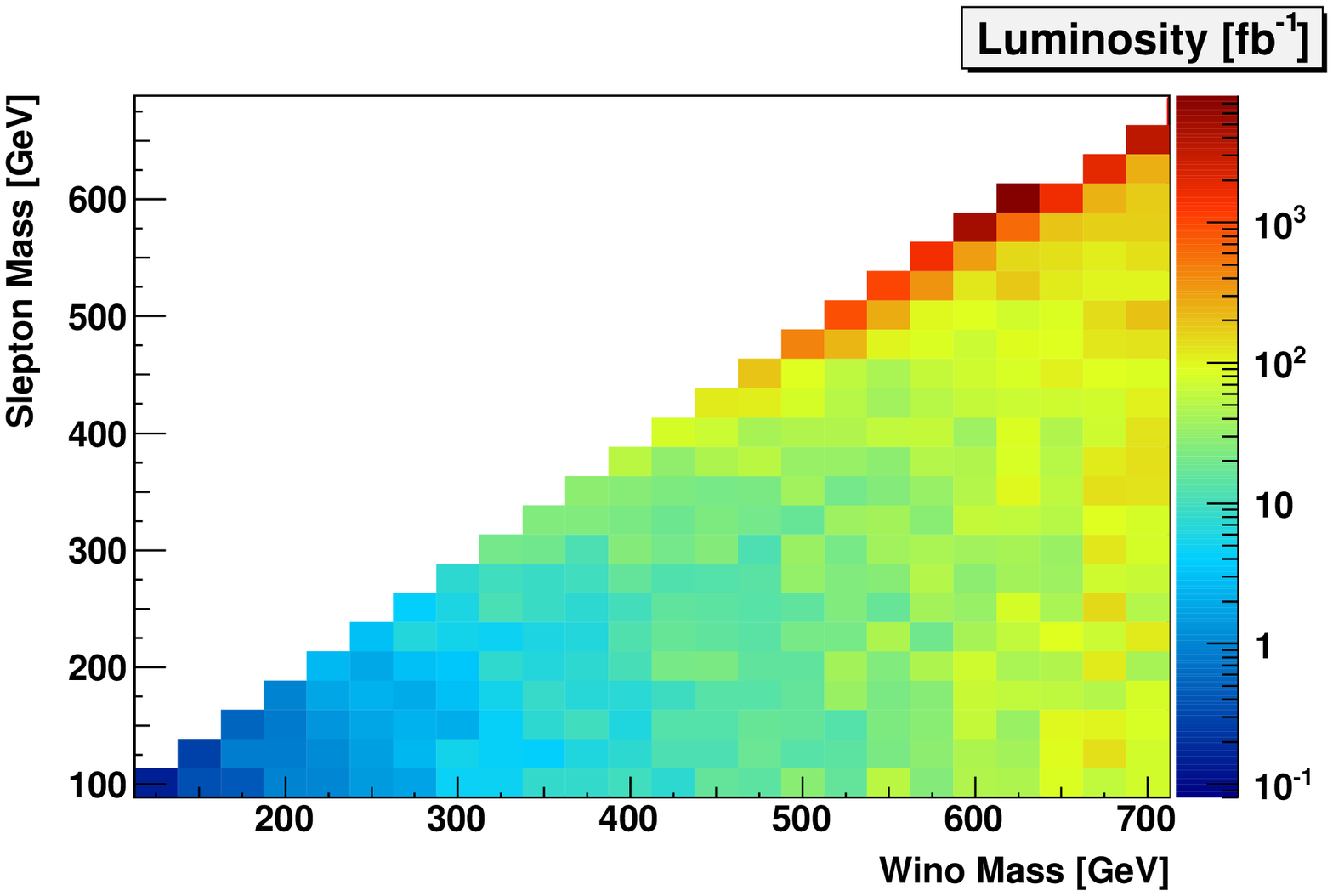}
\end{center}
\caption{\label{fig:lumi} Luminosity required at the 14 TeV LHC for a
  5$\sigma$ detection of the cutoff feature above the SM backgrounds.
  In the left plot, we assume only the left-handed selectron is accessible in the
  cascade decay of $\chi_2^0$.  In the right plot, we assume that all slepton masses are degenerate.  The cross section reaches are obtained considering
  the final states with electrons and muons  only.  }
  \label{fig:lum_reach}
\end{figure}

Wino pair production $\chi_1^\pm \chi_2^0$ is completely controlled by gauge couplings in this
approximation, which allows us to make robust predictions for the cross section
of this process. 
 With our fitting strategy, for a given mass parameter set with known
production cross section, we allow the luminosity to shift until
the triangle spectral shape in $m_{\ell\ell}$ distribution is detectable at the $5\sigma$ level, as
defined above.  The previous limits from LEP II constrain the
left-handed sleptons to be heavier than about 100 GeV, and so we consider only sleptons
which are 100 GeV or heavier. While charginos are more weakly
constrained in general than the left-handed slepton, they must be heavier than
the left-handed slepton in order to be within this framework.  
We scan $m_{\tilde{\ell}_L}$ and $M_2$ in the range of 
100 $-$ 700 GeV, with a step size of 25 GeV.  
The luminosities necessary to discover the slepton in
these decays as a function of the masses are shown in
Fig.~\ref{fig:lumi}. 

In the left plot of Fig.~\ref{fig:lumi}, we
assume that only one lepton flavor (selectron in our analysis) is
accessible in the cascade decay of $\chi_2^0$.    For the 14 TeV LHC with 100 ${\rm fb}^{-1}$ luminosity, the mass
reach for the $\tilde{\ell}_L$ can be reached up to 600 GeV at 5$\sigma$ level.  In the right plot, we
assume the lepton universality condition such that
$m_{\tilde{e}_L}=m_{\tilde{\mu}_L}=m_{\tilde{\tau}_L}$.  Considering
only trilepton events  with electrons and muons, namely $eee$, $\mu\mu\mu$,
 $e^+e^-\mu^\pm$ and $\mu^+\mu^- e^\pm$;
 additional branching ratio
suppression factors  need to be applied.   
The reach for the left-handed slepton mass is reduced in this case, about 
500 GeV for $m_{\tilde{\ell}_L}$.    

For comparison, Drell-Yan slepton searches were considered
in Ref.~\cite{Bityukov}, with the authors concluding that flavor-diagonal
sleptons with masses less than $350-400$ GeV  can be discovered by CMS
using 100 ${\rm fb}^{-1}$ of data.   
Thus, our technique indicates that, in
the pure gaugino limit, the reach for left-handed sleptons will be
significantly enhanced by looking for the characteristic cutoff shape of $m_{\ell\ell}$ in the
decays of the gauginos. Note that right-handed sleptons are not subject to
these results because they are a singlet under ${\rm SU}(2)_L$.

\section{Conclusions and Discussion}
\label{sec:conclusion}

In this work, we have studied the LHC discovery potential for  sleptons,
 which are produced via the on-shell decay of the heavier neutralino and chargino states.
 In particular, we have studied the  $\chi_1^\pm\chi_2^0$ associated production,
 with the consequent decays of $\chi_1^\pm\rightarrow   \nu_\ell \ell \chi_1^0$ and $\chi_2^0
 \rightarrow \ell \ell \chi_1^0$ via on-shell sleptons.
 Comparing to the conventional slepton searches through Drell-Yan production
 and dilepton plus missing $E_T$ final states, this trilepton 
plus missing $E_T$ channel has the advantage of larger production
 cross sections and less SM backgrounds.  In addition, the invariant mass
 distribution of the dilepton pair from $\chi_2^0$ decay has a distinctive triangle shape,
 which can be utilized to select out the signals from the dominant
 SM background of $WZ/\gamma^*$ and $t\bar{t}$ fakes. We performed
 a fit to the $m_{\ell\ell}$ distribution of both the signal and 
 the backgrounds.  For the LHC with 14 TeV center of mass energy and
 100 ${\rm fb}^{-1}$ integrated luminosity, we obtained the
 model-independent lower bounds on the effective signal cross section,
 $\sigma \times {\rm BR} \times {\rm acceptance}$,  as
 a function of triangle cutoff mass, $m_{\rm cut}$, at 5$\sigma$ significance level.
 Applying this result to the MSSM in the parameter space of $M_1<m_{\tilde\ell_L}<M_2 \ll \mu$,
 we found that  the mass
reach for the $\tilde{\ell}_L$ can be up to 600 GeV at 5$\sigma$ level
 at the 14 TeV LHC with ${\cal L}=$100 
 ${\rm fb}^{-1}$, when there is only one slepton
 generation (selectron in our study) lighter than Winos.
 For a degenerate slepton spectrum
 with $m_{\tilde{e}_L}=m_{\tilde{\mu}_L}=m_{\tilde{\tau}_L}$, and considering
 final states with electrons and muons only, the reach is slightly worse due
 to the suppression of the branching fractions.

Comparing to earlier studies on the LHC reach for sleptons from Drell-Yan production,  the reach for left-handed sleptons via Wino decay is greatly enhanced.  On the other hand, it should be noted that our study   works most effectively for the left-handed slepton, since the decay fraction of heavier  neutralino/chargino states to right-handed sleptons is typically suppressed in most of parameter space, either by the small Bino-Wino/Higgsino mixing, or the lepton Yukawa couplings.   The right-handed slepton could appear in Bino-like neutralino decay, if it is not the LSP.  The associated production cross sections for Bino with other neutralino/chargino states, however, are suppressed in general.     Our study does not apply to the parameter region of $m_{\tilde\ell_L} > M_2$ since the on-shell decay of   Winos into left-handed sleptons is forbidden by kinematics.  Therefore, for left-handed sleptons with mass heavier than $M_2$ or for right-handed sleptons, the usual Drell-Yan production is still the dominant production mode.

It should also be emphasized that the results we obtained for the lower bounds
 on the effective signal cross section, $\sigma \times {\rm BR} \times {\rm acceptance}$,
 as a function of triangle cutoff mass scale $m_{\rm cut}$ is model independent
 since it can be applied to other new physics models that give rise to the same cascade decay
 topology and final states of trilepton plus missing $E_T$.  For new physics models with a 
 given parameter set, we can obtain $m_{\rm cut}$ as well as the production cross sections,
 branching fractions into the trilepton final states, and signal acceptance via detector simulation.
 Comparing it with the lower bounds we obtained, we can derive the LHC reach in the parameter space 
 for such new physics models. 

More work is needed to fully explore the LHC reach for the slepton sector.  
Our results on the left-handed slepton reach are obtained under the simple assumption 
that $\chi_1^0$ is a pure Bino state, $\chi_2^0, \chi_1^\pm$ are pure Wino states, 
and  the heavier Higgsino states are completely decoupled.  In addition,  we studied only   final states including   electrons and muons.  The analysis strategy in our study can be applied to the  general MSSM framework with the mixing of gaugino and Higgsino states, as well as three lepton flavors (and possible left-right mixing in the stau case).  The corresponding branching fraction into trilepton final states needs to be taken into account in such general cases.  Trilepton final states with all possible flavor combinations could also be studied,   although the fitting to the triangle shapes might be more complicated when the slepton masses are not degenerate. 
 
In our study, we performed the triangle shape fitting to the trilepton plus missing $E_T$ final states.  The triangle shape in $m_{\ell\ell}$ distribution arises from the $\chi_2^0$ cascade decay chain.  Therefore, it appears in any   process that contains  such a heavy neutralino cascade decay.  The fitting strategy could be applied to   final states containing two opposite sign same flavor dileptons.  For example,  charginos in $\chi_1^\pm\chi_2^0$ production could decay to jets instead of leptons; or we could consider productions originated from gluinos or squarks, with the cascade decay of gluino or quark containing a $\chi_2^0$.
 SM backgrounds for dilepton plus jets plus missing $E_T$ signature, of course, are   very different from the trilepton plus missing $E_T$ signal that we considered in our study.

We could also consider Wino type $\chi_1^+\chi_1^-$ production with $\chi_1^\pm \rightarrow \nu_\ell \ell \chi_1^0$.   The final state of dilepton plus missing $E_T$ is similar to the conventional slepton study of Drell-Yan pair production of $\tilde\ell_L \tilde\ell_L$ with $\tilde\ell_L \rightarrow \ell \chi_1^0$.  Although this channel is not as powerful as the trilepton plus missing $E_T$ study that we explored in this paper, it would still have better reach comparing to the Drell-Yan process given the larger production cross sections.

As we mentioned earlier, our analyses can not be applied to the cases when the sleptons are heavier than $M_2$, when the Drell-Yan is the dominant production mode.  Previous LHC analyses on the Drell-Yan channel focused on the dilepton plus missing $E_T$ final states.  When  left-handed sleptons are heavier than Wino-like neutralino/charginos,  the branching fractions of heavier sleptons into Wino-like states are sizable given its ${\rm SU}(2)$ coupling strength.  Considering the consequent decay of the Wino-like neutralino/charginos, multiple leptons (up to six) + jets $+\  {\rm missing}\ E_T$ final states could appear, which provides additional channels for the left-handed slepton discovery at the LHC. 
For the right-handed sleptons, however, decays into lepton plus Bino-like $\chi_1^0$ LSP are still   dominant.

Light sleptons could also contribute sizably to low energy processes, 
for example, parity-violating electron scattering, leptonic Pion and Kaon decays,
 etc. \cite{michael_su}.  Given the recent progress on both the theoretical
 and experimental studies in those low energy precision measurements, they have
 reached a sensitivity which is now able to probe   new physics beyond the SM.  LHC studies
 on the slepton sector will   be complementary  to the indirect probes provided by
 these precision measurements.

\section{Acknowledgments}

We would like to thank M. Ramsey-Musolf, T. Han and S. Padhi for useful discussions. 
JE and WS would like to thank TASI where some of the work was completed.
JE and SS are supported by the Department of Energy
under Grant~DE-FG02-04ER-41298.  SS is also supported 
by NSF Grants No. PHY-0653656 and 
PHY-0709742. WS is supported in part by NSF Grant No. PHY-0970171.



\begin{thebibliography}{99}



\bibitem{ATLAS}
%
  G.~Aad {\it et al.} [ ATLAS Collaboration ],
    [arXiv:1109.6572 [hep-ex]];
%
  [arXiv:1109.6606 [hep-ex]];
ATLAS-CONF-2011-039;
  [arXiv:1110.2299 [hep-ex]].

\bibitem{ATLAS2L}
  G.~Aad {\it et al.} [ ATLAS Collaboration ],
  [arXiv:1110.6189 [hep-ex]].
  

\bibitem{CMS}
CMS Collaboration,
CMS-PAS-SUS-11-010;
%
  CMS-PAS-SUS-11-017;
%
 CMS-PAS-SUS-11-015;
CMS-PAS-SUS-11-003;
%
CMS-PAS-SUS-11-004;
CMS-PAS-SUS-11-005.

\bibitem{CMSmultilep}
  CMS Collaboration,
 CMS-PAS-SUS-11-013.
 
 
 \bibitem{CMSOS2L}
CMS Collaboration,
CMS-PAS-SUS-11-011.

  
 

\bibitem{padhi}T.~Han, S.~Padhi and S.~Su, ``Searching for Charginos and Neutralinos at the LHC", in preparation.

\bibitem{GMSB} For a review, see 
  G.~F.~Giudice, R.~Rattazzi,
  Phys.\ Rept.\  {\bf 322}, 419-499 (1999) 
  [hep-ph/9801271].

\bibitem{AMSB} 
  L.~Randall, R.~Sundrum,
  Nucl.\ Phys.\  {\bf B557}, 79-118 (1999) 
  [hep-th/9810155];
  G.~F.~Giudice, M.~A.~Luty, H.~Murayama, R.~Rattazzi,
  JHEP {\bf 9812}, 027 (1998)
  [hep-ph/9810442];
  T.~Gherghetta, G.~F.~Giudice, J.~D.~Wells,
  Nucl.\ Phys.\  {\bf B559}, 27-47 (1999)
  [hep-ph/9904378].
  
  
  

\bibitem{mSUGRA} 
For a review, see 
  A.~B.~Lahanas, D.~V.~Nanopoulos,
  Phys.\ Rept.\  {\bf 145}, 1 (1987).

\bibitem{neutralinoDM} 
  H.~Goldberg,
  Phys.\ Rev.\ Lett.\  {\bf 50}, 1419 (1983);
  J.~R.~Ellis, J.~S.~Hagelin, D.~V.~Nanopoulos, K.~A.~Olive, M.~Srednicki,
  Nucl.\ Phys.\  {\bf B238}, 453-476 (1984).
  


\bibitem{sleptonrelic}M.~Drees, M.~M.~Nojiri,
  Phys.\ Rev.\  {\bf D47}, 376-408 (1993)
  [hep-ph/9207234];
  T.~Nihei, L.~Roszkowski, R.~Ruiz de Austri,
  JHEP {\bf 0203}, 031 (2002)
  [hep-ph/0202009];
  A.~Birkedal-Hansen, E.~-h.~Jeong,
  JHEP {\bf 0302}, 047 (2003)
  [hep-ph/0210041].
 

\bibitem{earlyslepton}

  F.~del Aguila, L.~Ametller,
  Phys.\ Lett.\  {\bf B261 } (1991)  326-333.
  
  

\bibitem{Baer}
  H.~Baer, C.~-h.~Chen, F.~Paige, X.~Tata,
  Phys.\ Rev.\  {\bf D49}, 3283-3290 (1994)
  [hep-ph/9311248];
  H.~Baer, C.~-h.~Chen, F.~Paige, X.~Tata,
  Phys.\ Rev.\  {\bf D53}, 6241-6264 (1996)
  [hep-ph/9512383].
  
  
  
\bibitem{Bityukov}
  S.~I.~Bityukov, N.~V.~Krasnikov,
  Phys.\ Atom.\ Nucl.\  {\bf 62}, 1213-1225 (1999)
  [hep-ph/9712358];
  Y.~.M.~Andreev, S.~I.~Bityukov, N.~V.~Krasnikov,
  Phys.\ Atom.\ Nucl.\  {\bf 68}, 340-347 (2005)
  [hep-ph/0402229].
D.~Denegri, L.~Rurua and N.~Stepanov, CMS note TN/96-059, 1996. 

\bibitem{Baergaugino}
  H.~Baer, K.~Hagiwara and X.~Tata,
  Phys.\ Rev.\  D {\bf 35}, 1598 (1987);
  H.~Baer and X.~Tata,
  Phys.\ Lett.\  B {\bf 155}, 278 (1985).
  
\bibitem{Hinchliffe:1996iu}
  I.~Hinchliffe, F.~E.~Paige, M.~D.~Shapiro, J.~Soderqvist, W.~Yao,
  Phys.\ Rev.\  {\bf D55}, 5520-5540 (1997)
  [hep-ph/9610544].
  
\bibitem{Birkedal:2005cm}
  A.~Birkedal, R.~C.~Group, K.~Matchev,
 [hep-ph/0507002].
  
   
\bibitem{Abdullin:2006jm}
  S.~Abdullin {\it et al.} [ TeV4LHC Working Group Collaboration ],
   [hep-ph/0608322].
  
 
\bibitem{sleptonflavor} 
J.~L.~Feng, S.~T.~French, I.~Galon, C.~G.~Lester, Y.~Nir, Y.~Shadmi, D.~Sanford, F.~Yu,
  JHEP {\bf 1001}, 047 (2010)
  [arXiv:0910.1618 [hep-ph]];
  J.~L.~Feng, S.~T.~French, C.~G.~Lester, Y.~Nir, Y.~Shadmi,
  Phys.\ Rev.\  {\bf D80}, 114004 (2009)
  [arXiv:0906.4215 [hep-ph]].
  
\bibitem{Krasnikov:1996np}
  N.~V.~Krasnikov,
  JETP Lett.\  {\bf 65}, 148-153 (1997)
  [hep-ph/9611282].
  S.~I.~Bityukov, N.~V.~Krasnikov,
    [hep-ph/9806504].

 

 \bibitem{LEPslepton}
 LEP2 SUSY Working Group,  ``Combined LEP Selectron/Smuon/Stau Results, 183-208 GeV", (2004),
 note LEPSUSYWG/04-01.1, 
 http://lepsusy.web.cern.ch/lepsusy/.
 
\bibitem{LEPsinglelep}
  P.~Achard {\it et al.} [ L3 Collaboration ],
  Phys.\ Lett.\  {\bf B580}, 37-49 (2004)
  [hep-ex/0310007].
  A.~Heister {\it et al.} [ ALEPH Collaboration ],
  Phys.\ Lett.\  {\bf B544}, 73-88 (2002)
  [hep-ex/0207056].  
  
\bibitem{:2005ema}
  [ ALEPH and DELPHI and L3 and OPAL and SLD and LEP Electroweak Working Group and SLD Electroweak Group and SLD Heavy Flavour Group Collaborations ],
  Phys.\ Rept.\  {\bf 427}, 257-454 (2006)
  [hep-ex/0509008].


 \bibitem{LEPchargino1}LEP SUSY Working Group, LEPSUSYWG/01-03.1.

\bibitem{LEPchargino2}  LEP2 SUSY Working Group, LEPSUSYWG/02-04.1.
 
 \bibitem{LEPLSP}
 LEP2 SUSY Working Group, LEPSUSYWG/04-07.1.
 
 
  \bibitem{LEPLSPmSUGRA}
 LEP2 SUSY Working Group, LEPSUSYWG/02-06.2.
 
 
  
   \bibitem{CDFtrilepton}
  T.~Aaltonen {\it et al.} [ CDF Collaboration ],
  Phys.\ Rev.\ Lett.\  {\bf 101}, 251801 (2008).
  [arXiv:0808.2446 [hep-ex]];
  CDF Collaboration, CDF note 10636.
 
\bibitem{Abazov:2009zi}
  V.~M.~Abazov {\it et al.} [ D0 Collaboration ],
  Phys.\ Lett.\  {\bf B680}, 34-43 (2009)
  [arXiv:0901.0646 [hep-ex]].

  
  

\bibitem{Sullivan:2008ki}
  Z.~Sullivan, E.~L.~Berger,
  Phys.\ Rev.\  {\bf D78}, 034030 (2008)
  [arXiv:0805.3720 [hep-ph]].


\bibitem{Rajaraman:2010hy}
  A.~Rajaraman, F.~Yu,
  Phys.\ Lett.\  {\bf B700}, 126-132 (2011)
  [arXiv:1009.2751 [hep-ph]].

  


\bibitem{Alwall:2011uj}
  J.~Alwall, M.~Herquet, F.~Maltoni, O.~Mattelaer and T.~Stelzer,
  JHEP {\bf 1106}, 128 (2011)
  [arXiv:1106.0522 [hep-ph]].
  
   \bibitem{MadGraph}
  J.~Alwall {\it et al.},
  JHEP {\bf 0709}, 028 (2007)
  [arXiv:0706.2334 [hep-ph]].
  
  
  \bibitem{PYTHIA}
  T.~Sjostrand, S.~Mrenna and P.~Skands,
  JHEP {\bf 0605}, 026 (2006)
  [arXiv:hep-ph/0603175].

 \bibitem{PGS}
PGS -- Pretty Good Simulator,
http://www.physics.ucdavis.edu/$\sim$conway/research/software/
pgs/pgs4-general.html.

  
\bibitem{MINUIT}
  F.~James and M.~Roos,
  Comput.\ Phys.\ Commun.\  {\bf 10}, 343 (1975).
   
\bibitem{michael_su}
A.~Kurylov, M.~J.~Ramsey-Musolf and S.~Su,
  Phys.\ Rev.\  D {\bf 68}, 035008 (2003)
  [arXiv:hep-ph/0303026];
  Phys.\ Lett.\  B {\bf 582}, 222 (2004)
  [arXiv:hep-ph/0307270];
M.~J.~Ramsey-Musolf, S.~Su and S.~Tulin,
  Phys.\ Rev.\  D {\bf 76}, 095017 (2007)
  [arXiv:0705.0028 [hep-ph]];
For a review, see M.~J.~Ramsey-Musolf and S.~Su,
  Phys.\ Rept.\  {\bf 456}, 1 (2008)
  [arXiv:hep-ph/0612057].

  
  
  
  
  
  
 
 
 
 
  
  
  
  
  

  
  

   
 
  


\end{thebibliography}
\end{document}